\documentclass[journal]{vgtc}                % final (journal style)
\ifpdf%                                % if we use pdflatex
  \pdfoutput=1\relax                   % create PDFs from pdfLaTeX
  \usepackage{graphicx}                % allow us to embed graphics files
  \DeclareGraphicsExtensions{.pdf,.png,.jpg,.jpeg} % for pdflatex we expect .pdf, .png, or .jpg files
\else%                                 % else we use pure latex
  \usepackage{graphicx}                % allow us to embed graphics files
  \DeclareGraphicsExtensions{.eps}     % for pure latex we expect eps files
\fi%

\graphicspath{{figures/}{pictures/}{images/}{./}} % where to search for the images
\usepackage{color}
\usepackage{microtype}                 % use micro-typography (slightly more compact, better to read)
\PassOptionsToPackage{warn}{textcomp}  % to address font issues with \textrightarrow
\usepackage{textcomp}                  % use better special symbols
\usepackage{mathptmx}                  % use matching math font
\usepackage{times}                     % we use Times as the main font
\usepackage{cite}                      % needed to automatically sort the references
\usepackage{tabu}                      % only used for the table example
\usepackage{booktabs}                  % only used for the table example
\usepackage[ruled,vlined]{algorithm2e}
\usepackage{amssymb}
\usepackage{fontawesome}
\usepackage{dirtytalk}
\usepackage{tikz}
\usepackage{tabu}  
\usepackage{tabularx}
\usetikzlibrary{shapes}
\usepackage{makecell}
\usepackage{multicol}
\usepackage{multirow}
\usepackage{colortbl}
\usepackage[most]{tcolorbox}
\usepackage[framemethod=TikZ]{mdframed}
\usepackage{lipsum}% http://ctan.org/pkg/lipsum
\usepackage{stfloats}
\usepackage{titlesec}
\usetikzlibrary{plotmarks}
\newcommand{\tikzcircle}[2][red,fill=red]{\tikz[baseline=-0.5ex]\fill[#1,radius=#2] (0,0) circle ;}%
\usepackage{enumitem}

\newcommand{\sysname}{\textit{SDR}Querier}
\newcommand{\newdesign}{Temporal Availability Profiler}
\definecolor{darkgreen}{rgb}{0.0, 0.7, 0.0}

\newcommand{\clrg}{\textcolor{darkgreen}}

\onlineid{1169}

\vgtccategory{Research}
\vgtcpapertype{application/design study}

\title{\sysname{}: A Visual Querying Framework for\\ Cross-National Survey Data Recycling}

\author{Yamei Tu, Olga Li, Junpeng Wang, Han-Wei Shen, Przemek Powalko, \\ Irina Tomescu-Dubrow, Kazimierz M. Slomczynski, Spyros Blanas and J. Craig Jenkins}
\authorfooter{
\item
 Yamei Tu, Han-Wei Shen, Spyros Blanas, J.Craig Jenkins are with The Ohio State University. E-mail: tu.253, shen.94, blanas.2, jenkins.12 @osu.edu.
\item
Junpeng Wang is with Visa Research. E-mail: junpeng@vt.edu.
\item
 Olga Li, Przemek Powalko, Irina Tomescu-Dubrow, Kazimierz M. Slomczynski are with Institute of Philosophy and Sociology. Email: li.olya.en@gmail.com, powal@yahoo.com, dubrow.4, slomczynski.1 @osu.edu
}

\shortauthortitle{Biv \MakeLowercase{\textit{et al.}}: Global Illumination for Fun and Profit}
\abstract{
Public opinion surveys constitute a widespread, powerful tool to study peoples’ attitudes and behaviors in comparative perspectives. 
However, even world-wide surveys provide only partial geographic and time coverage, which hinders a comprehensive knowledge production. 
To broaden the scope of comparison, social scientists turn to \textit{ex-post} harmonization  
 of variables from datasets that cover similar topics but in different populations and/or years. 
The resulting new datasets can be analyzed as a single source, which can be flexibly accessed through many data portals for scientists. However, such portals offer little guidance to explore the data in-depth or query data with user customized needs.
As a result, it is still challenging for social scientists to efficiently identify related data for their studies and evaluate their theoretical models based on the sliced data.
To overcome these limitations, in the Survey Data Recycling (SDR) international cooperation research project, we propose \sysname{} and apply it to the harmonized SDR database, which features over two million respondents interviewed in a total of 1,721 national surveys that are part of 22 well-known international projects spanning the period 1966-2012 and 142 countries/territories. Using the SDR database as a prototype, we design the \sysname{}
to solve three practical challenges that social scientists routinely face. First, a BERT-based model provides customized data queries through research questions or keywords (\textit{Query-by-Question}). 
Second, we propose a new visual design to showcase the availability of the harmonized data at different levels, thus helping users decide if empirical data exist to address a given research question (\textit{Query-by-Condition}). Lastly, \sysname{} discloses the underlying relational patterns among substantive and methodological variables in the database (\textit{Query-by-Relation}), to help social scientists rigorously evaluate or even improve their regression models. Through case studies with multiple social scientists in solving their daily challenges, we demonstrated the novelty, usefulness and effectiveness of \sysname{}.}

\CCScatlist{ % not used in journal version
 \CCScat{K.6.1}{Management of Computing and Information Systems}%
{Project and People Management}{Life Cycle};
 \CCScat{K.7.m}{The Computing Profession}{Miscellaneous}{Ethics}
}

\teaser{
  \centering
  \includegraphics[width=\linewidth]{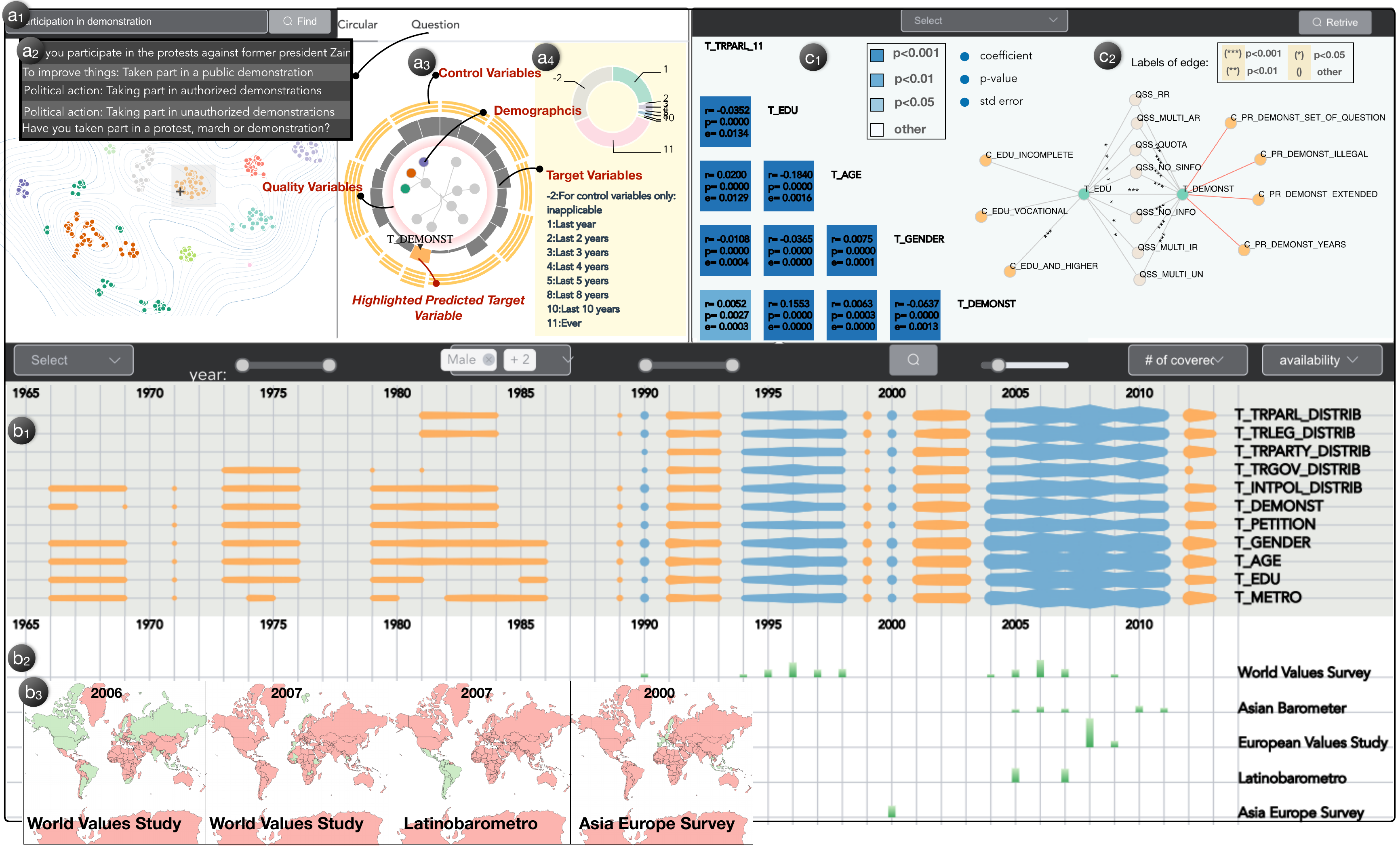}
  \vspace{-0.25in}
  \vspace{-0.06in}
  \caption{(a1-a4) \textit{Query-by-Question} recommends target variables based on user input and allows further visual explorations; (b1-b3) \textit{Query-by-Condition} includes the new design, \newdesign{}, revealing the multi-faceted information of available data given user's conditions; (c1-c2) \textit{Query-by-Relation} presents the relational patterns from the available data to assist social research. }
  \label{fig:teaser}
}

\vgtcinsertpkg

\begin{document}


% \firstsection{Introduction}

\maketitle

%% \section{Introduction} %for journal use above \firstsection{..} instead

\section{Introduction}
Comparative surveys are a powerful tool that researchers in many fields, such as sociology, political science, economics, demography, etc., employ to study how the individual-level conditions (e.g., age, gender) combine with contextual factors (e.g., democracy, economics) to shape social phenomena across cultures and time~\cite{krosnick1999survey,singleton2009approaches}.  
While a treasure of free, publicly accessible international survey projects exists, users encounter drawbacks in doing comparative analyses, mainly because single survey projects cover only a fraction of the world's nations and selected time periods.
To broaden the scope of comparison, social scientists increasingly harmonize information from existing cross-national datasets that measure the same concepts for different populations and/or years into a new integrated  database~\cite{ruggles2003ipums,frick2007cross}.
Survey Data Recycling (SDR) is such an active research project that develops ex-post harmonization methods \cite{granda2010harmonizing,tomescu2014democratic} to recode, rescale, or transform variables from 22 international surveys into one integrated dataset with consistent scales ~\cite{kolczynska2018item,oleksiyenko2018identification,SDQ2021}. 
The SDR harmonized database is available online through the SDR data portal, such that the scientists can flexibly access the data to conduct further analysis.
The large-scale harmonized databases that entail the potential for innovative comparative research are also likely to raise substantial difficulties in \textit{understanding} and 
\textit{exploring} the dataset, as well as \textit{evaluating} their theoretical models built on top of the sliced data. Given the organization of the current online data portal, scholars generally do not have access to effective means for \textit{understanding} the complex structure and various types of variables, causing difficulties in choosing an appropriate 
set of variables from the data for their analyses. It is also difficult for researchers to \textit{explore} data availability taking into account source data quality or harmonization features for their analysis, even with a set of accurate filtering conditions. 
Lastly, survey data are used to \textit{evaluate} regression models proposed by scientists. Nevertheless, retrieving the useful information from the available high-quality data to evaluate the fit of statistical models against the empirical data is a non-trivial task.

With the success of visualization in analyzing multi-variate and multi-faceted data, we believe it is key to solving the above challenges from three aspects. First, as SDR data are harmonized from a set of meta-data, i.e. survey questionnaires, codebooks, and data dictionaries, illustrating structures of the harmonized data and relating the unstructured texts with the harmonized variables can significantly improve the effectiveness of data query. 
Second, as both the meta-data and harmonized data suffer from severe data quality issues, demonstrating data availability is in strong need, which avoids spending time on less-verifiable problems but initiate promising research topics with solid data support.
Third, visualizations with convenient user interactions can greatly assist in the exploration of the hidden relationships between the meta-data and harmonized data in the dataset.
Apart from these potential benefits, however, we found the power of visualizations has not been sufficiently leveraged in social science applications. 
For example, bar charts are adopted frequently to show the temporal coverage of surveys, but they fall short in revealing the surveys' spatial coverage simultaneously. Scatterplots are commonly used to qualitatively present the correlation between target variables, but they fail to reflect the quality of the underlying data and may present biased results \cite{healy2014data}.

To overcome these limitations, we collaborate with social scientists and use the SDR database as a pilot case for developing a new visual analytical system, the \sysname{}.
The system is equipped with three-level of information queries through visualization and interactions. 
To facilitate \textit{understanding}, we propose a question-driven variable recommendation for efficient data exploration. Users can query relevant variables by inputting their research questions.
Based on the related variables, users can perform accurate queries to check available data. 
For \textit{exploring} data availability, 
% In order to present availability for decision making,
\sysname{} is equipped with a new design, \textit{\newdesign{}}, which exhibits multi-faceted information dynamically. 
Furthermore, our system performs model evaluation and suggests methodological variable improvements by answering the following questions: \textit{What are the relationships between variables selected for the regression model?} and \textit{What other variables are necessary to include in the model?} 
We also perform extensive case studies and host thorough discussions with domain experts. 
To sum up, the main contributions of our work are as follows:
\begin{itemize}[leftmargin=0.18in, topsep=-0.6em]
	\itemsep-0.25em
    \item We abstract the challenges in analyzing harmonization survey data and propose \textbf{a visual analytics system}, \sysname{}, to solve them. It is equipped with three visual components assisting in different stages: understanding, exploring, and analyzing. 
    \item We propose a new \textbf{question-driven variable recommendation} for data understanding, which facilitates users to identify variables of interest in an efficient way. %It is also confirmed to be useful in other scenarios by our domain experts. 
    \item We design \textbf{\newdesign{}} to visualize available survey projects from different levels and perspectives. We prove the novelty and usefulness of this design with thorough case studies.
\end{itemize}
\section{Related Work}

\subsection{Survey Data Visualization}
To present information succinctly, social scientists frequently use  basic visualizations to generate static reports \cite{healy2014data}. For example, for discrete categorical
data, bar charts or pie charts are commonly utilized to display proportions or distributions of different categories ~\cite{jones2016web }. 
For quantitative data, bar charts, such as boxplots and error bars, are designed to incorporate statistical measurements  ~\cite{jones2016web,ryssevik2001social}. 
These static visualizations can only convey information formed and filtered by the creators of visualization. To allow human-in-the-loop of the information seeking process, there are some visualization tools that allow users to flexibly explore survey data with their own questions, e.g., NESSTAR~\cite{assini2002nesstar}, SDA~\cite{thomas2011sda}. 
Jones et al. developed an interactive system for presenting quantitative social environmental survey data to help explore and understand~\cite{jones2016web}. 
All the works mentioned above aim to understand the content of survey data through visualization. To our knowledge, 
\sysname{} is the first interactive system that allows users to explore large, complex and high-dimensional harmonized dataset through visualization and multi-faceted queries.

\subsection{Time-Varying Multivariate Data Visualization}
Time varying multivariate data depict how various features evolve over time, and these evolution patterns often provide valuable insights into the data generated by different domains~\cite{aigner2007visualizing,wongsuphasawat2011lifeflow}. 
Based on the fact that time can be considered either linear or cyclic, visualizations can be categorized into two groups: time series plots~\cite{liu2012tiara,luo2010eventriver} vs. spiral graphs~\cite{hewagamage1999interactive,carlis1998interactive,weber2001visualizing}. The Spiral Graph is more efficient to discern periodic patterns. For sequential visualization, the Theme River~\cite{havre2000themeriver}, is one of the most popular visualization that maps the frequencies of multiple topics at each time step to the widths of colored currents in the river, depicting the thematic evolution of documents based on a river metaphor. 
In our \newdesign{}, the sub component, \textit{Separate Availability} uses the same metaphor as the Theme River, presenting the data availability as a flow. However, we allow the information presented from multiple perspectives, where Theme River and other methods are not applicable~\cite{harris1999information,hewagamage1999interactive}. 
There are many visualization tools designed for capturing multi-level information of time-varying data in the literature~\cite{fails2006visual}. 
Dasgupta et al. developed coordinated views to illustrates the evolution of chemical species for geologists to observe interactions, including parallel coordinates and matrix views~\cite{dasgupta2015vimtex}.
Wang et al. designed a spiral graph for analyzing sentiment of time-varying twitter data~\cite{wang2015senticompass}.
Pena et al. compared three visualizations of geo-temporal multivariate data, which is considered as the most related work~\cite{pena2019comparison} to this paper. However, the difference is that geological and temporal information fall into two levels of our analysis in social research. The purpose of \newdesign{} is to illustrate the temporal availability and high-level spatial availability first, and the detailed geological information is presented later to scientists.

\subsection{BERT for Information Retrieval}
Information Retrieval has advanced rapidly in the recent years due to the development of natural language processing (NLP) technology. Given the state-of-the-art performance in many downstream tasks in NLP, the most related ones to our automatic recommendation model are search-related, such as document retrieval and question answering~\cite{dibia2020neuralqa,nogueira2019passage,yang2019end}.  
Some works applied BERT to ad-hoc document retrieval by ranking the documents based on inference scores computed for each document given a specific query~\cite{yang2019simple,macavaney2019cedr,yilmaz2019applying}. 
However, they differ in the way of computing the inference score. Yang el al. 
tackled the challenge of long documents by inferring individual sentence first and then computing document score based on the sentences~\cite{yang2019simple}, 
while Jiang et al. handled cross-lingual document retrieval between English queries and foreign-language documents~\cite{jiang2020cross}.  
The aforementioned works aim to improve the performance or solve the difficulties when apply BERT to information retrieval, while for \sysname{}, we propose to use BERT for variable recommendation from two perspectives. Also, we identify the different scenarios to apply this model. 

\vspace{-0.3em}
\section{Background} \label{SEC:BACKGROUND}
In this section, we introduce the different types of variables in the harmonized data, meanwhile explaining where the data complexity comes from. We also provide a brief introduction of the BERT model and its inner structure, used in \sysname{}.
\setlength{\belowcaptionskip}{-12pt}
\begin{figure}
    \centering
    \includegraphics[width=\linewidth]{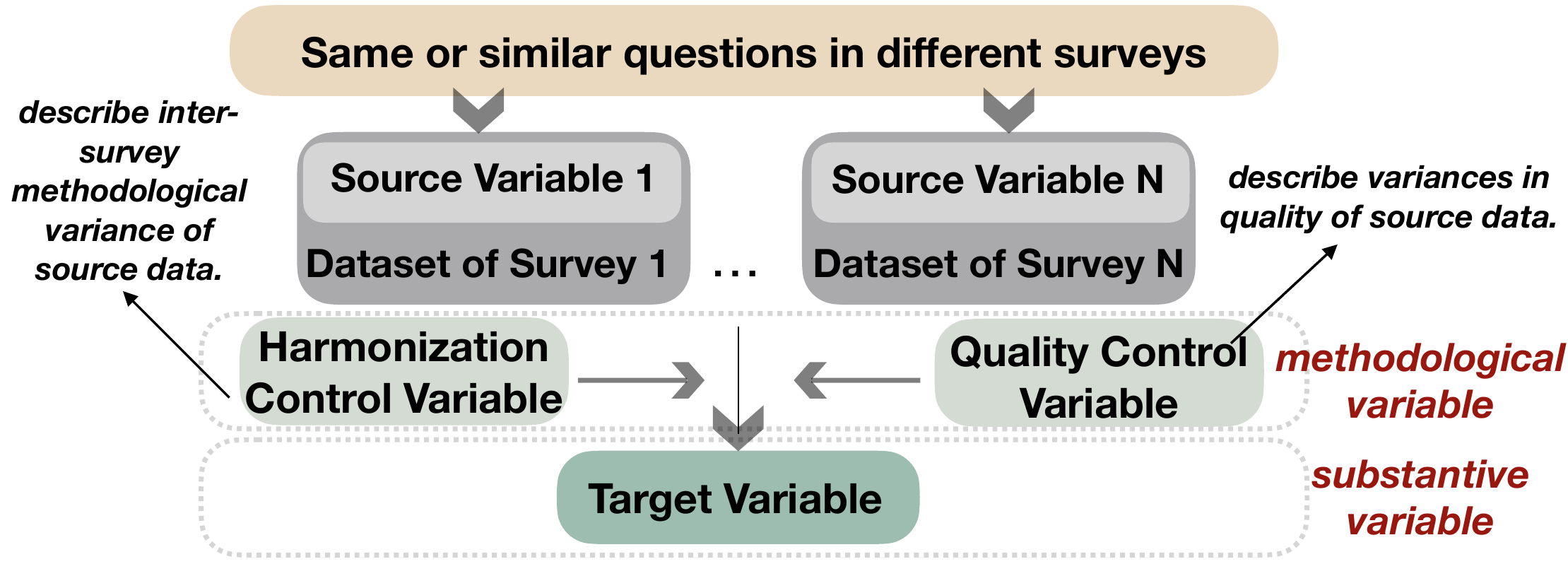}
    \vspace{-0.28in}
    \caption{The relationships of different variables in the harmonized data.}
    \label{fig:FIGURE1}
\end{figure}
\setlength{\belowcaptionskip}{0pt}
\subsection{Variables Types in Harmonized SDR Dataset}
During ex-post survey data harmonization, information from international survey projects and other sources is combined into the integrated SDR dataset. Given one survey, each question in the questionnaire is related to one dimension (column) in the dataset. For example, a survey may ask the respondents questions about their attitudes and behaviors and record all responses into a table. In the resulting tabular dataset, each column corresponds to a specific question, and each row indicates one respondent’s answers to the questions in the survey questionnaire. Each dimension refers to one \textit{variable}. Original variables, taken from different surveys for harmonization, are called \textit{source variables}. The indicator in the harmonized dataset, produced from a series of source variables measuring the same concept in different surveys, is called a \textit{target variable}.
Combining the information from different source variables into a target variable requires ex-post harmonization procedures, since the characteristics (e.g. wording, answer options) of source questions related to one concept frequently vary between surveys.

During the process of transforming source variables into target variables, the SDR team creates \textit {harmonization controls}. These are target variable specific measures that capture primarily inter-survey methodological variability in formulation of the source questions that can influence the validity and reliability of the constructed target variable.

The SDR database provides another set of methodological indicators, source data \textit {quality controls}. These variables capture biases and errors that stem from differences in the quality of the source survey data, where quality is operationalized along three dimensions: the source documentation (questionnaires, codebooks, study descriptions, technical reports), data records in individual source datasets, and the consistency between these documentation and data records. Source data quality controls also can affect the relationship between substantive target variables. Researchers should assess if and to what extent they do so. 
Overall, ex-post survey data harmonization in the SDR project yields the following types of variables: 
\begin{itemize}
[leftmargin=0.15in, topsep=-0.6em]
	\itemsep-0.25em 
    \item \textbf{Source:} raw variables from the surveys taken for harmonization.
    \item \textbf{Target:} substantive variables in the integrated dataset, constructed out of source variables as the product of ex-post harmonization. 
    \item \textbf{Harmonization Control:}  methodological variables that accompany target variables to record properties of the source variables that can affect the reliability and validity of the target variable, and that could be lost in the process of transforming source into target variables.
    \item \textbf{Quality Control:} methodological variables that address inter-survey variations and the quality of the source survey data.  
\end{itemize}

\subsection{BERT}
 In this section, we first introduce the BERT model, then move on to describe the basic component inside the model, namely, Encoder.  
 
 \textbf{BERT Model:} Bidirectional Encoder Representations from Transformers (BERT) is a Transformer-based language representation model that can be fine-tuned to achieve state-of-the-art performance on many natural language processing tasks~\cite{devlin2018bert}. 
It origins from pre-training contextual representations, e.g. ELMo~\cite{peters2018deep},  ULM-FiT~\cite{howard2018universal}, OpenAI~\cite{radford2018improving}, etc. 
The BERT coverts an input sequence $(x_{1},...,x_{n})$ to a sequence of vector representations $\textbf{z}=(z_{1},...,z_{n})$~\cite{vaswani2017attention}. 
The BERT outperforms previous work by considering the context for each occurrence of one word, which means BERT generates different embeddings for the same words in different contexts.
As shown in \autoref{fig:FIGURE3}(A), BERT is composed of a stack of identical Transformer Encoders. There are two model sizes: $BERT_{BASE}$(Encoder$\times$12) and $BERT_{LARGE}$(Encoder$\times$24). We use $BERT_{BASE}$ in the automatic recommendation model.

\textbf{Transformer Encoder.} The encoder of BERT is based on the original implementation of Transformer~\cite{vaswani2017attention}. As shown in \autoref{fig:FIGURE3}(B), it has two sub-layers: the multi-head attention layer and the feed-forward network. 
In the first layer, for head i, it first multiplies the input embedding matrix with three learnable parameter matrices $W_{i}^{Q}$, $W_{i}^{K}$,$W_{i}^{V}$ into Q, K, V and generates the output matrix as: 
\setlength\abovedisplayskip{1pt}
\setlength\belowdisplayskip{1.5pt}
\begin{equation}\small
    Z_{i} = Attention(Q_{i},K_{i},V_{i}) = softmax(\frac{Q_{i}K_{i}^{T}}{\sqrt{d_{k}}})V_{i}
\end{equation}
In $BERT_{BASE}$, there are 12 heads, which means 12 sets of $(Q_{i},K_{i},V_{i})$ attending on different information. 
So the output of the multi-head attention layer is then calculated as:
\begin{equation}\small
    MultiHead(Q,K,V) = Concat(Z_{1},...,Z_{12})W^{O}
\end{equation}
 where~ $W^O\in \mathbb{R}^{12\times d_{v}\times d_{model}}$. 
The second layer consists of two linear transformations with a ReLU activation in between:
\begin{equation}\small
    FFN(x) = max(0, xW_{1}+b_{1})W_{2}+b_{2}
\end{equation}
\section{Requirement Analysis and Approach Overview} \label{SEC:requirement}
\subsection{Design Requirements}
The SDR portal enables scientists to download harmonized  survey data that can be used for comparative empirical research. The challenge is how to help them identify what the related data are and how to use them. We have collaborated with four domain experts for more than one year to identify the requirements,  summarized as follows:
% Two experts have 45+ years experience in cross-national research, one expert has 20+ years experience in survey data harmonization and another expert has  5+ years experience in both survey data harmonization and data management. 
\begin{itemize}[leftmargin=0.18in, topsep=-0.6em]
	\itemsep-0.25em
    \item \textbf{R1: Identifying related variables to the user's research topic.} Given the large dimensionality of the harmonized dataset, identifying the related columns is important and necessary to acquire meaningful data from the portal. In order to provide enough guidance for experts, \sysname{} is required to:
    \begin{itemize}[leftmargin=0.18in, topsep=-0.6em]
	\itemsep-0.25em
        \item \textit{R1.1: Give the variable recommendation based on users' needs.} Automatic variable recommendation can help scientists avoid unnecessary exploration and focus  on more important variables. 
        \item \textit{R1.2: Exhibit data provenance of harmonized target variables.} Showing what source variables each target variable links to helps scientists understand the logic and meaning of each target variable. Simultaneously, this background information fosters researchers' trust in the harmonized data, as it speaks to the transparency of the harmonization process. 
        \item \textit{R1.3: Present an overview of the harmonized dataset.} Due to the complexity of the harmonized data, plenty of information need to be presented (e.g., types of variables, relations between variables) for understanding these data. 
        
        % such as the types of variables, the relationships among variables, and the distribution and labels of different values for each variable. 
    \end{itemize}
    \item \textbf{R2: Revealing data availability for decision support.} Typically, scholars who conduct quantitative comparative survey research seek data that meet specific conditions. Retrieving valid records by specific conditions is an easy task, but deciding whether they are sufficient to evaluate a scientific model should take many other factors into consideration.

% , such as diversity, quality, temporal variability, and spatial coverage, etc.
    % For example, "trust in institutions affects the protest behavior for US females with different education and living conditions" restricts the country and gender of respondents. 
    \begin{itemize}[leftmargin=0.18in, topsep=-0.6em]
	\itemsep-0.25em
        \item \textit{R2.1: Facilitating target variables selection.} While several target variables may fit to a given research problem, their availability varies a lot.  In order to decide which one to choose, scientists need to know their individual and joint availability.
        \item \textit{R2.2: Assisting with decision making.} Once available data are identified, it is important to assist researchers in deciding whether these data meet the formal requirements for regression analysis. This can be done by providing multi-faceted information, e.g., which data have quality issues?
    \end{itemize}
    \item \textbf{R3: Retrieving underlying patterns for hypothesis testing.} Social scientists use survey data to examine if and to what extent there is empirical support  hypotheses between various variables, which can be assisted by the  hidden patterns from the data.
    \begin{itemize}[leftmargin=0.18in, topsep=-0.6em]
	\itemsep-0.25em
         \item \textit{R3.1 Validating the selected target variables.} Hypotheses propose some associations or causal relationships between variables of interest. Revealing relational patterns from target data is a good way to preliminarily evaluate the hypotheses.
        \item \textit{R3.2 Describing the potentially related variables to improve the regression model.} Relations between target variables should also consider the potential role of methodological variables. Scientists should take them into consideration when building theoretical models to test hypotheses.
        % Given the novelty of the SDR data, these relations cannot be learned in literature reviews, requiring \sysname{} to retrieve this information for social scientists to build models. }
        % For example, when studying the protest behavior, scientist may propose a theory that more educated respondents are more likely to protest,  and build a regression model based on the hypothesis. However, not all available data can be fit into the theory. How to identify the existence of supportive data is important for model evaluation. 
          
    \end{itemize}
\end{itemize}

\subsection{Approach Overview}
\setlength{\belowcaptionskip}{-12pt}
\begin{figure}
    \centering
    \includegraphics[width=\linewidth]{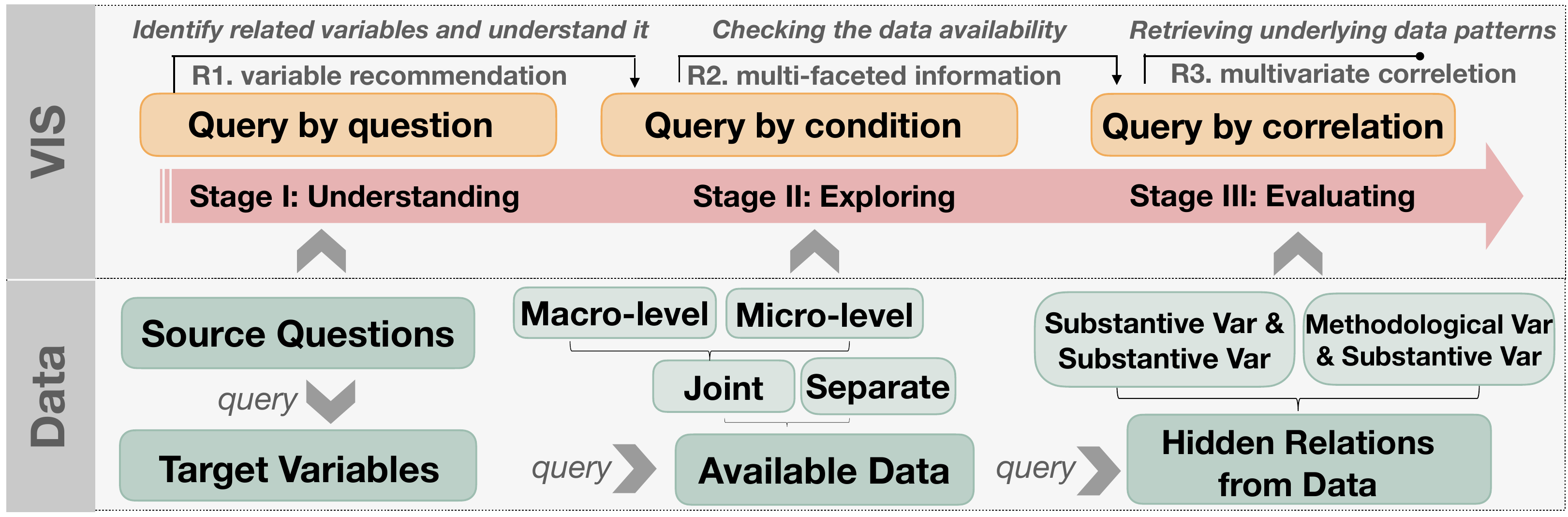}
    \vspace{-0.25in}
    \caption{The overview of our framework to benefit harmonized data analysis in social research.}
    \label{fig:FIGURE2}
\end{figure}
\setlength{\belowcaptionskip}{0pt}

\autoref{fig:FIGURE2} displays an overview of our framework. 
We summarize the domain requirements into three challenges in different stages of the social research pipeline: \textit{understanding, exploring and evaluating}. 
To solve the challenges, we propose a framework that contains three corresponding modules.
\textbf{First}, inspired by conversational Artificial Intelligence, we train a BERT-based model to generate variable recommendations based on user's input text, either keywords or sentences describing their information of interest. This process is
defined as \textit{Query-by-Question}. 
Later, the recommendation is combined with visualization and interactions to facilitate harmonized data understanding.
\textbf{Second}, in the \textit{Query-by-Condition} module, we perform information retrieval based on specific filtering conditions. In order to show the multi-faceted information from the retrieved data, we design a new visualization, \newdesign{}, to assist scientists in deciding whether data are sufficient to use considering data diversity, coverage, and quality issues. 
\textbf{Lastly}, computing the relational patterns from available data samples can verify whether the expected patterns exist or not, which in turn helps scientists to test their hypotheses and choose the variables for their theoretical models, defined as \textit{Query-by-Relation.} 

\section{Visual Analytics System: \sysname{}}
Motivated by the requirements in \autoref{SEC:requirement}, we design and implement \sysname{} with three coordinated visual components, enabling multi-granularity queries of the harmonized survey data for social scientists.

\subsection{Query-by-Question (QBQ)}
\label{SEC:QUESTION}
Although target variables names in the harmonized data have been carefully chosen, it can be difficult  to quickly identify the theoretical concept from the abbreviated names. 
As explained in \autoref{SEC:BACKGROUND}, each target variable is summarized from a set of survey questions in the questionnaires. Therefore, the survey questions provide good contexts, accurately reflecting the meaning of target variables.
For example, \texttt{T\_DEMONST} (a target variable) can be characterized as \textit{authorized demonstrations in democratic countries} or \textit{unauthorized  activities in non-democratic countries} given different political backgrounds.

Inspired by conversational AI, we train a BERT-based classification model on survey questions to predict the target variables. 
With such a model, we can infer the target variable from a wide variety of text inputs, e.g., research questions, descriptions, or a set of keywords for a sociological concept. For example, when a researcher studies if life conditions can influence political participation, he/she might type in the sociological concept, i.e.\say{political participation}, or the descriptions of life condition indicators, i.e.\say{how much are you satisfied with your life?} or \say{are you living in metropolitan or not?} to retrieve the related target variables.
Based on the model, we can recommend a target variable in two ways (\textbf{R1.1}): (1) \textit{the hard recommendation}, which outputs the target variable with the highest probability from the classification model; (2) \textit{the soft recommendation}, which converts one-to-one prediction problem to a one-to-many clustering issue by allowing users to flexibly explore the semantic similarity between their inputs and survey questions. 
% \clrb{which 
% extracts the hidden states, i.e. embeddings of the input texts from the trained model and allows users to flexibly explore it in the context of questions' embeddings.}

\subsubsection{BERT-Based Model for Target Variable Prediction} \label{SEC:BERT-PREDICTION}
To automate the QBQ process, we train a BERT-based model to relate the survey questions with target variables. The model (1) takes a survey question as input, (2) embeds it into a [cls] token, which represents the entire text sequence and is then used for sequence classification tasks, and (3) converts the [cls] token to a target variable. A pre-trained BERT model is employed to perform (1)${\rightarrow}$(2), and a classification layer is appended to the model to conduct (2)${\rightarrow}$(3). A set of question and target variable pairs, labeled by our social scientists, are used to train the classification layer with a cross-entropy loss. 

% We have questions which has been labeled with their target variables manually by domain experts, which is used as training and validation dataset. 
% Our model is a pre-trained BERT with one classification layer. 
% We feed the questions into the BERT, the output embedding of [cls] token represent the entire sentence, which is used for classification tasks.
% The classification layer calculate the possibility of assigning [cls] token embedding to each class.
% We use cross-entropy as loss function, defined as:
% \begin{equation}
%     H(p,q) = -\sum_{x}p(x)logq(x)
% \end{equation}where p(x) and q(x) represent predicted probability and actual probability respectively. 
% The computed loss is then back-propagated to update the parameters. 
% We use batch-size=32, epochs=4, lr = 2e-5, eps = 1e-8 in our training. The loss and accuracy details can be seen from \autoref{}\clr{TBA}.
\setlength{\belowcaptionskip}{-12pt}
\begin{figure}
    \centering
    \includegraphics[width=\linewidth]{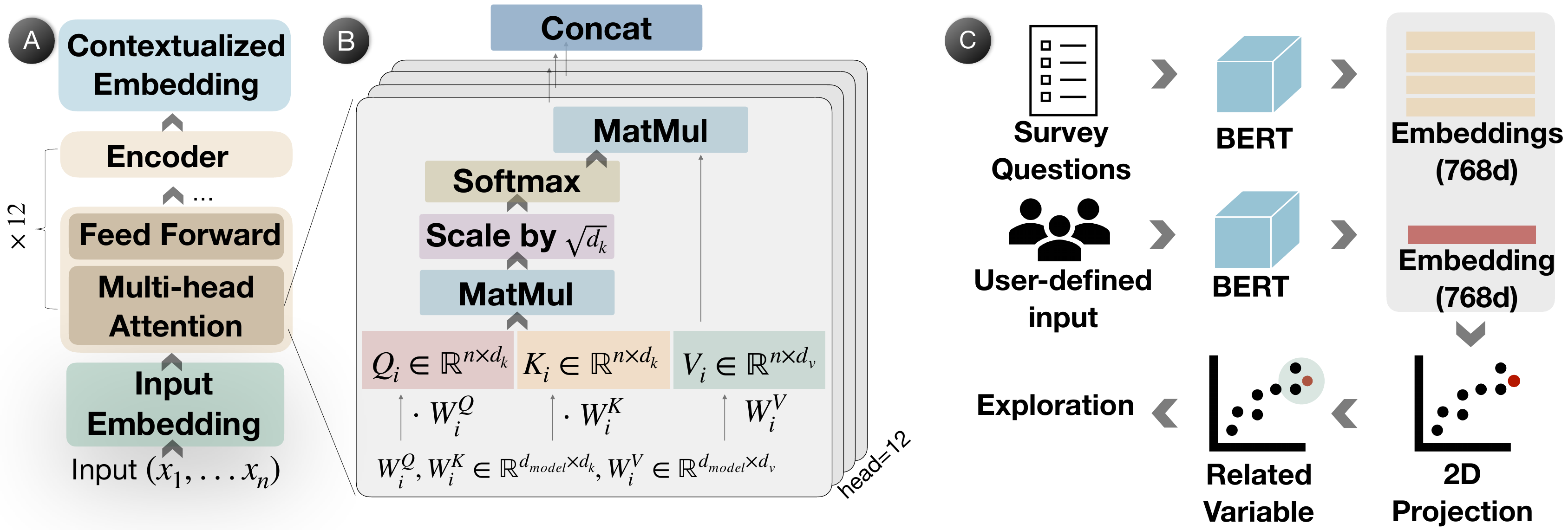}
    \vspace{-0.25in}
    \caption{(A) Structure of $BERT_{BASE}$; (B) illustration of the Transformer Encoders; (C) pipeline of BERT-based Variable Recommendation;}
    \label{fig:FIGURE3}
\end{figure}
\setlength{\belowcaptionskip}{0pt}

% This component displays three types of relationships between variables: source vs. source, source vs. target, target vs. target. 
% Without a deep and clear understanding of these relationships, it is impossible for social scientists to perform accurate query to the harmonized survey data. 
% In this component, we guide scientists to the target variables by querying their defined research questions through text semantic similarity, defined as \textit{Query by Question}. In this way, it can help scientists to define the sub columns when they want to query data from portal.
% After locating to the information of interest, we also provide the dynamic explorations and rich interactions to facilitate details analysis. 

\subsubsection{BERT-Based Model for Soft Recommendation}
% 1. goal of the method 
The \textit{soft recommendation} qualitatively measures the semantic similarity between user-defined text input and the survey questions.
We extract the hidden states, i.e. embeddings from the trained model, which is promised to capture the semantic information. 
As shown in \autoref{fig:FIGURE3}(C), the embeddings of users' input and survey questions are extracted and jointly projected to 2D for visual exploration. tSNE~\cite{van2008visualizing} is employed here to interactively update the projection result, given its superior performance over UMAP for non-linear projections. We perform the embedding updates in an iterative manner following \autoref{alg: BERT}, aiming to reduce the running time and acquire stable results.
% After generating embeddings, we apply dimensionality reduction algorithm, i.e. UMAP~\cite{mcinnes2018umap}, to project the embeddings into 2D space. We perform the updates in an iterative manner following \autoref{alg: BERT}. There are several advantages over non-iterative updating. First, the coordinates of question are tending to be stable during the updating process. Second, the convergence of UMAP algorithm is faster initializing with last timestamp compared to random initialization.  

\setlength{\intextsep}{0pt}% Remove \textfloatsep
\begin{algorithm}[h]
\SetAlgoLined
\KwIn{question projection coordinates at timestamp t, $\mathbb{P}^{t}$: ($p_{1}^{t}$,...,$p_{N}^{t}$), new input sentence: s}
\KwOut{whole projection coordinates set $\mathbb{P}$ at timestamp t+1, $\mathbb{P}^{t+1}$: ($p_{1}^{t+1}$,...,$p_{N}^{t+1}$, $p_{s}^{t+1}$)}
\caption{{\bf    Embedding Iterative Updating Algorithm} \label{alg: BERT}}

\nl $e_{s}^{t+1}$ = BERT(s) \clrg{//Generating embedding for new input s}

\nl $p_{s}^{t+1}$ = random\_init($e_{s}^{t+1}$) \clrg{//Random initializing position for s}

\nl $\mathbb{P}^{t}$ = ($p_{1}^{t}$,...,$p_{N}^{t}$, $p_{s}^{t+1}$) \clrg{//Adding coordinate of s into $\mathbb{P}$}

\nl $\mathbb{P}^{t+1}$ = tSNE(init=$\mathbb{P}^{t}$) \clrg{//Init tSNE with $\mathbb{P}$ from last timestamp}

\end{algorithm}
% As a result, the similar questions will be closed to each other and form cluster\autoref{fig:FIGURE3}(D).  
% The \textit{soft recommendation} process received positive feedback from the domain experts and inspired them to identify similar questions in their on-going harmonization works. We will quantitatively evaluate the  result from \textit{soft recommendation} in the evaluation part. 

% After the discussion with experts, we find that this method is also helpful to identify similar questions to harmonize in harmonization pre-processing. We will evaluate the clustering result with ground truth labels in the evaluation part.
% Based on projection and clustering, the relationships between target variables and user interested information can be converted to the 2D distance between questions and user-defined questions.
% Later, we allow scientists to brush the closed question embeddings to check the corresponding target variable and explore to acquire more detailed information. 

\subsubsection{Visual Design for Understanding Harmonized data}
% 1. the role of three visual component and their relationship, 
There are three coordinated views to assist the identification of related data in the harmonized database (\textbf{R1}): the \textit{Scatterplot} in  \autoref{fig:teaser}-$a_{1}$, the \textit{Information Table} in \autoref{fig:teaser}-$a_{2}$, and the \textit{Circular Graph} in  \autoref{fig:teaser} $a_{3}$-$a_{4}$.

% \clr{what is the ``targeted source variables'' in the following paragraph? I understand this is an overview but i feel it can be deleted for a succinct description. Your system section already took a lot space now.}
% The scatterplot presents the relationships of source variables (i.e., questions) and user-defined input to help scientists identify targeted source variables \textbf{(R1.1)}.
% The detailed information of the targeted source variables are updated in the tabular information, bridging the transition from source variable to related variable \textbf{(R1.2)}. 
% Once users understand how different target variables are related to their focused research question, they can dive into the circular graph to see the relationship among target variables, along with the detailed information from the harmonized data \textbf{(R1.3)}. 

% 2. embedding projection, scatterplot
The \textit{Scatterplot} demonstrates the embedding projection result, revealing semantic similarity among survey questions and the user-defined input. 
Each dot represents one question, relating to one target variable, so we use the target variable to color the questions in the projection space.
As shown in \autoref{fig:teaser}(a1), questions of the same color are grouped together, verifying that our BERT model captures their semantic similarity.
It also presents the variance in the same source variables.
% For example, in \autoref{fig:FIGURE3}(D1), a user types in one research question: ``life satisfaction, protest behavior''. We can see that this information is projected between the green dot (``Satisfaction with your life'') and the blue dot (``Have you taken part in a protest demonstration in the last twelve months? (interviewer: exclude demonstrations against terrorism)''). 
%It confirms that the text with similar semantic meanings will be closed to each other in the projection space.  \clr{update your example as you mentioned in the comments}
From the projection, users can also brush the dots of interest, which will update the \textit{Information Table} automatically. 

% 3. tabular information 
The \textit{Information Table} connects source information and target information together, aiming to help scientists identify the data columns to query from the the SDR portal. The columns of the tabular data are \textit{year}, \textit{survey wave}, \textit{source question}, \textit{target variable}, \textit{label of target variable}. As confirmed by domain experts, individual source question varies across surveys, and hence it is helpful to present this variation to scientists in order to help them better understand the data pre-processing process  and improve the credibility of the harmonized data \textbf{(R1.2)}.

% 4. circular graph
The \textit{Circular Graph} is proposed to handle the complexity and dimensionality of the harmonized data, which targets to: (1) indicate diverse types of variables, such as \textit{source-, target-,  harmonization control- or quality-variables}; (2) illustrate the relationships of different variables (\textbf{R1.3}). As shown in \autoref{fig:teaser}(a3), the circular bar chart represents the target variables. The length of the bar implies the overall availability of each target variable, i.e., how frequently the corresponding target variable is measured in international surveys.
The color indicates the topic of the target variables, which is consistent with the color schema used in the \textit{Scatterplot}.
Once the user triggers the query from \textit{scatterplot}, only the predicted bar will be highlighted in orange while others fade out. 
Several target variables can describe the same topic from different perspectives. For example, \texttt{T\_HAPPY\_11} and \texttt{T\_HAPPY\_DISTRIB} both measure respondents' self-reported happiness, but using different specifications.  
As described in \autoref{SEC:BACKGROUND}, target variables capturing the same theoretical concept can share  one or several harmonization control variables, which record the variance in source variable properties. These controls are visualized in the orange arcs (\tikzcircle[fill={rgb,255:red,255; green,180; blue,0}]{3pt}) , the number of arcs in the same radial position reflects the number of harmonization control variables in the group. When clicking an arc, the right panel will pop up to show the value distribution and value label of the harmonization control variable(\autoref{fig:teaser}-$a_{4}$). As proved by experts, knowing the meaning and distribution of harmonization control variables is extremely helpful when querying data from the SDR portal. 
The inner circle (\tikzcircle[fill={rgb,255:red,253; green,224; blue,221}]{3pt}) conveys that all the target variables are related to the \textit{quality control} variables.
The most inner network represents the information and demographics of the respondents. The demographics include respondents' \textit{age}, \textit{birth year}, \textit{sex} of the respondents; their color is also consistent with the \textit{Scatterplot}. For example, for \textit{age}, survey can ask about age in many ways as reflected by the numerous red points (\tikzcircle[fill={rgb,255:red,217; green,95; blue,2}]{3pt}) in the  \textit{Scatterplot}.  

% \vspace{-0.8em}
\subsection{Query-by-Condition (QBC)}\label{SEC:AVAIL}
% Sociological hypotheses are proposed based on literature studies or distilled from related publications and are tested with the appropriate data samples.
Core to social science quantitative comparative research is to assess the extent to which empirical data provide support to their hypotheses. 
While these postulated hypotheses often refer to specific countries and certain year-range, it is a common practice among social scientists to blindly download the full harmonized data without any filtering conditions (from data portals), and then check if downloaded data fit their research needs, e.g., in terms of country and time coverage.
% capture enough variations; cover enough consecutive year; contain enough target variables necessary to conduct a comparative analysis.
However, the process is inefficient and can be greatly improved if 
data availability can be effectively and user-friendly checked from multiple perspectives before downloading \textbf{(R2)}.

\subsubsection{\newdesign{}}
% The \textit{Availability} view is designed to facilitate the interactive exploration of data availability in multiple levels. 

We propose a new design, \textit{\newdesign{}} to reveal the availability of the harmonized data at multiple levels. 
The design is composed of two sub-components (Separate Availability and Join Availability), sharing the same $x$-axis to reveal data samples density evolution across time (i.e., temporal availability).

The \textit{Separate Availability} view (\autoref{fig:teaser}-$b_{1}$) illustrates the amount of valid samples for \textbf{each} target variable over time, which helps user decide among multiple alternative variables \textbf{(R2.1)}.
The \textit{Joint Availability} view (\autoref{fig:teaser}-$b_{2}$) exhibits the available samples for \textbf{all} the selected target variables, indicating the precise pool of valid samples that one can rely on to  evaluation the multi-variate relations \textbf{(R2.2)}.
\setlength{\belowcaptionskip}{-12pt}
\begin{figure}
    \centering
    \includegraphics[width=\linewidth]{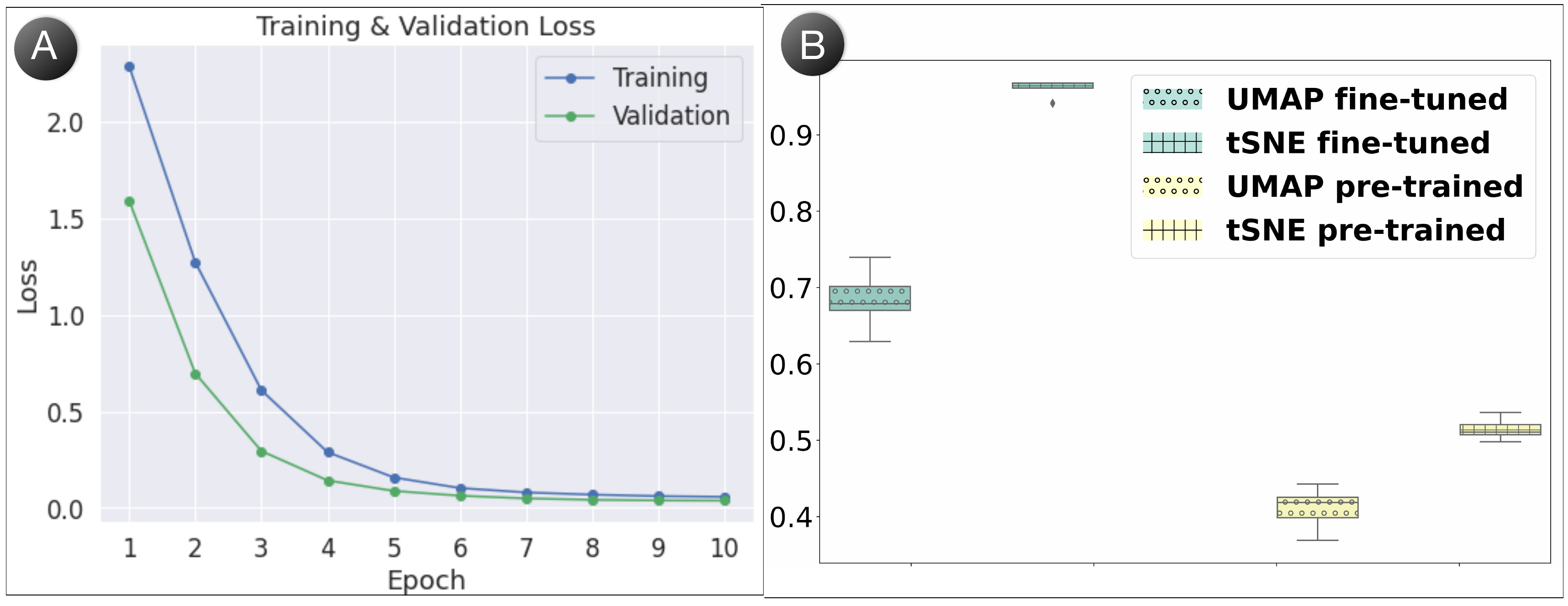}
    \vspace{-0.25in}
    \caption{(A) The training \& validation Loss of the BERT-based model. (B) Adjusted Mutual Information(AMI) score for the  pre-trained and fine-tuned models with different projections, i.e. UMAP and tSNE.}
    \label{fig:loss}
\end{figure}
\setlength{\belowcaptionskip}{0pt}

% The data generation and view construction is described in \autoref{alg: AVAILABILITY}. 
Before we construct the view, we have condition sets $\mathbb{C}$ and selected target variable sets $\mathbb{T}$. Condition sets are used to filter rows in the harmonized dataset. For example, a scientist wants to study political protests in Russia under Putin’s regime. The condition sets $\mathbb{C}$ = ("country=Russia", "year$\leq$2020", "year$\geq$2000"). 
For \textit{Joint Availability} view, the available samples should satisfy all condition sets, and also contain data at all target columns. 
% The samples group by different variables for developing multiple functions further \autoref{SEC:\newdesign{}}. 
While in the \textit{Separate Availability}, each row indicates one corresponding target variable $t_{j}$. The samples in each row should be valid for both condition sets and corresponding column.

Given one specific year, the connection between the two sub components can be summarized in the following situations:
\begin{enumerate}[leftmargin=0.18in, topsep=-0.6em]
	\itemsep-0.25em
    \item case1:  Each target variable $t_{i}$ has data $d_{i}$, and there are also jointly available samples, i.e., $\forall t_{i} \in \mathbb{T}, d_{i}\neq \varnothing \rightarrow d_{1}\cap d_{2}\cap... d_{N}\neq \varnothing$. This is the ideal case where there exist data of high-quality to use. 
    \item case2:  At least one target variable $t_{i}$ does not have data, resulting in the lack of jointly available data, i.e., $\exists t_{i} \in \mathbb{T}, d_{i} = \varnothing \rightarrow d_{1}\cap d_{2}\cap... d_{N}= \varnothing$. In other words, the lack of available data to use comes from specific variables, helping scientists to decide whether to omit the unavailable variable or impute the missing data. 
    \item case3:  Each target variable $t_{i}$ has data $d_{i}$, but there is no overlap among them, i.e., $\forall t_{i} \in \mathbb{T}, d_{i}\neq \varnothing \rightarrow d_{1}\cap d_{2}\cap... d_{N}=\varnothing$. This scenario indicates we have low-quality data since they do not contain all the variables of interest. 
\end{enumerate}

\vspace{-0.3em}
\subsubsection{Visual Design of \newdesign{}} \label{SEC:\newdesign{}}
As visualized in \autoref{fig:teaser}-$b_{1}$, \textit{Separate Availability} presents the available data for each variable over time. The color summarizes aforementioned connections with \textit{Joint Availability}, including blue (case1) and orange (case2 \& case3).
For each variable, the width of flow illustrates how many samples are covered each year. In the \textit{Joint Availability}, each row represents one valid survey project, which may cover a period of years. 
A user may click the survey project name to show the background information of each survey, incorporating survey documentation into \sysname{} is highly recommended by domain experts (\autoref{fig:Russian}C-D).
Given one survey, there is multi-faceted information to be presented properly. 
Through the discussion with domain experts, they prefer simple but efficient visualization to delicate glyph-design for multi-faceted information. 
To do this, we propose some interactions with a responsive bar chart to show information from multiple perspectives. 

\textbf{Responsive Bar Chart} The meaning of bar can be embedded as either \textit{macro-level} or \textit{micro-level} by users, defined as the \textit{responsive bar chart}. Scientists can select to see how many respondents are available \textit{(micro-level)} or how many countries are available \textit{(macro-level)} through the drop down selector \faChevronDown~ in the top of this module.
To present the country coverage, we further allow users to click the bar for the detailed information in a map, where green means covered country (\autoref{fig:teaser}-$b_{3}$).
Also, we allow two different sorting methods of rows: availability-based and quality-based. For the availability-based method, if a project covers more distinct  years, it will have higher availability. For quality-based sorting, we compute a quality indicator $q_{i}$ for each survey project (each row) as follows: 
\begin{equation}\small
    quality_{i} = \frac{\sum_{w_{i} \in \mathbb{S}}{N_{\sigma(q=0)~ in ~w_{i}}}}{\sum_{w_{i} \in \mathbb{S}}{N_{\sigma (\varnothing)~ in~ w_{i}}}}
\end{equation} Suppose one survey set $\mathbb{S}$ is composed of many waves $w_{i}$ conducted in different years. In each wave, there are some samples that do not have quality issues, which are recorded as quality variables $q=0$ in the harmonized dataset. The quality indicator is the faction of the samples without quality issues to the total samples given survey.  
% Separate Availability presents the availability for each target variable\textit{(row3)}. Each data satisfies condition sets and containing corresponding target variable\textit{(row2)}. 
% As shown in \clr{ref}, each row corresponds to one target variable $t_{j}$. At each time step, the width of the flow indicates how many samples are available for this variable and satisfying all the condition sets\textit{(row2)}. It can also be regard as a summary of joint availability. Since we sum the data from joint availability in each year, 
\subsection{Query-by-Relation (QBR)} \label{SEC:RELATION}

Social scientists propose hypotheses derived from the literature review, which are then tested through statistic models using appropriate data. 
However, building these models often requires a long time to process the survey data and identify the related variables. 
Effective and accurate variable selections becomes crucial.
% requiring a correct selection of variables when constructing the model. 

To accommodate this, we propose a third module, \textit{Query-by-Relation (QBR)}, to query the hidden patterns from data for model verification and improvements. QBR is performed by answering two questions: (1) \textit{what are the relationships between the selected target variables?} Whether the variables are correlated can help scientists to preliminary test their hypotheses. For example, a researcher wants to test whether respondent's resources have a negative effect on individual's trust in political institutions. Checking the correlation strength among these variables can help scientists to determine whether the selected variables are appropriate for hypotheses testing. (2) \textit{what are the potentially related variables?} In the SDR harmonized dataset, both types of methodological variables can affect the relationship between substantive variables, so they should be included in the regression analysis when constructing regression models. Understanding the correlation between them is important for scientists to include appropriate additional methodological variables into their models. To the end, we designed two subviews in QBR to answer the above two questions.
% Based on the two steps, we develop this QBR view, consisting of two subviews as follows.

\textbf{Correlation Matrix.} Driven by \textbf{R3.1}, our first subview presents the pairwise correlations for user-selected target variables, allowing scientists to check if these variables are correlated with each other and to further determine what to keep for their regression analysis. 
We compute several necessary and common statistics for pairwise relations: (1) Pearson correlation coefficient, (2) p-value, (3) levels of the p-value (thresholds are suggested by experts), (4) standard errors.  
To flatten the learning curve of visual encoding, as suggested by our domain experts, we show the computed information with texts and only incorporate two visual channels in the matrix, i.e., position for pairwise relation and responsive color. 
Users are allowed to select one of the computed information and map it to the color interactively. 
After several key design iterations with the domain experts, we determined to show one-half of the symmetric matrix to reduce redundant information and avoid unnecessary interpretation of the position.

% for several reasons. First, for the symmetrical relations, showing half matrix is sufficient and able to retrieve all the related information with one-time position searching. 
% Second, keeping design simple and easy to understand for scientists.}

\textbf{Network Visualization.} Variable selection is extremely important when building  regression models to test research hypotheses. 
Due to the dissimilar structure of harmonized data to typical survey data, users may not know what methodological variables should be considered together with substantive variables for a robust model analysis. 
% For example, harmonization controls are created in the process of harmonizing source variables to generate target variables, capturing the important properties that worth preserving.
Thus QBR facilitates to query the complex relations given one pair of target variables (\textbf{R3.2}). 
As shown in \autoref{fig:teaser}-$c_{2}$, we apply the same color scheme to the type of nodes as QBQ: harmonization control (\tikzcircle[fill={rgb,255:red,255; green,196; blue,120}]{3pt}), quality control (\tikzcircle[fill={rgb,255:red,240; green,229; blue,216}]{3pt}), target (\tikzcircle[fill={rgb,255:red,117; green,207; blue,184}]{3pt}). We label the significance level for each edge in the network, which is defined by domain experts. If correlation coefficient is not defined for an edge, we highlight it with red. 
\setlength{\belowcaptionskip}{-12pt}
\begin{figure}
    \centering
    \includegraphics[width=\linewidth]{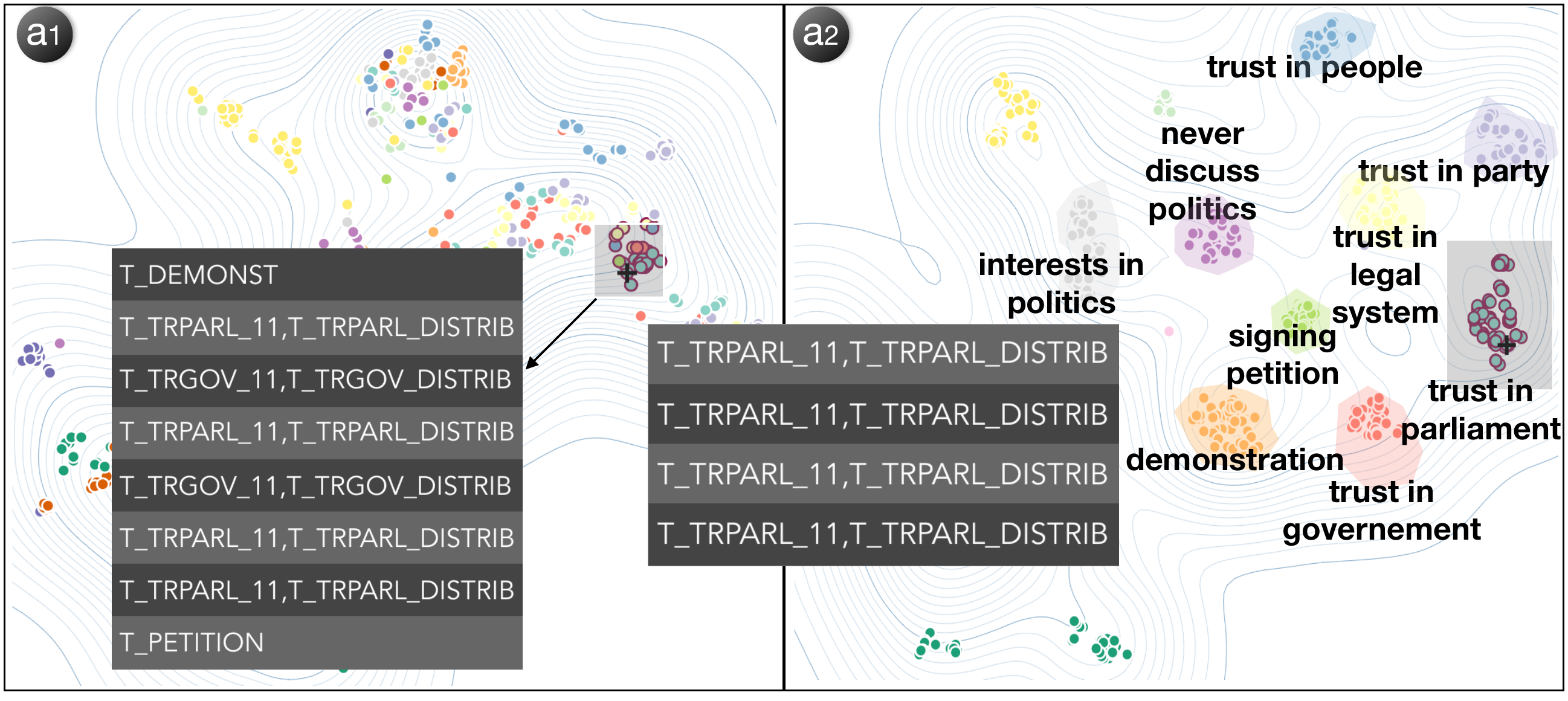}
    \vspace{-0.25in}
    \caption{ tSNE projection space of the (a1) pre-trained  and the (a2) fine-tuned  BERT model. }
    \label{fig:evalution_embedding}
\end{figure}
\setlength{\belowcaptionskip}{-12pt}

\section{Evaluation}

This section first evaluates the efficiency and usefulness of the backend algorithms employed in \sysname{}. 
Then, we worked with the domain experts on the case studies to demonstrate how \sysname{} can help them in the process of understanding and exploring harmonized data, as well as evaluating social science models via effectively visual-queries and friendly user-interactions. 
% Finally, we gather feedback of our system from four domain experts in sociology. 

\subsection{Evaluation of the BERT-based Model}
We fine-tune a pre-trained BERT model to classify each survey question to the most related target variable, aiming to recommend variables given multiple types of user-defined inputs. To achieve this, both the final prediction and intermediate result, i.e. embeddings are extracted from the tuned model for \textit{hard} and \textit{soft} recommendation respectively. To perform a comprehensive evaluation, we demonstrate the effectiveness of two recommendations through  qualitative and quantitative measures.
% The BERT-based model provides recommendations in two ways, which are evaluated differently. The efficiency of \textit{hard recommendation} is evaluated via training details and prediction results. 
% For \textit{soft recommendation}, we demonstrate effectiveness of the embedding space through both qualitative and quantitative measures.

\subsubsection{Quantitative Evaluation}
\setlength{\belowcaptionskip}{-12pt}
\begin{figure}
    \centering
    \includegraphics[width=\linewidth]{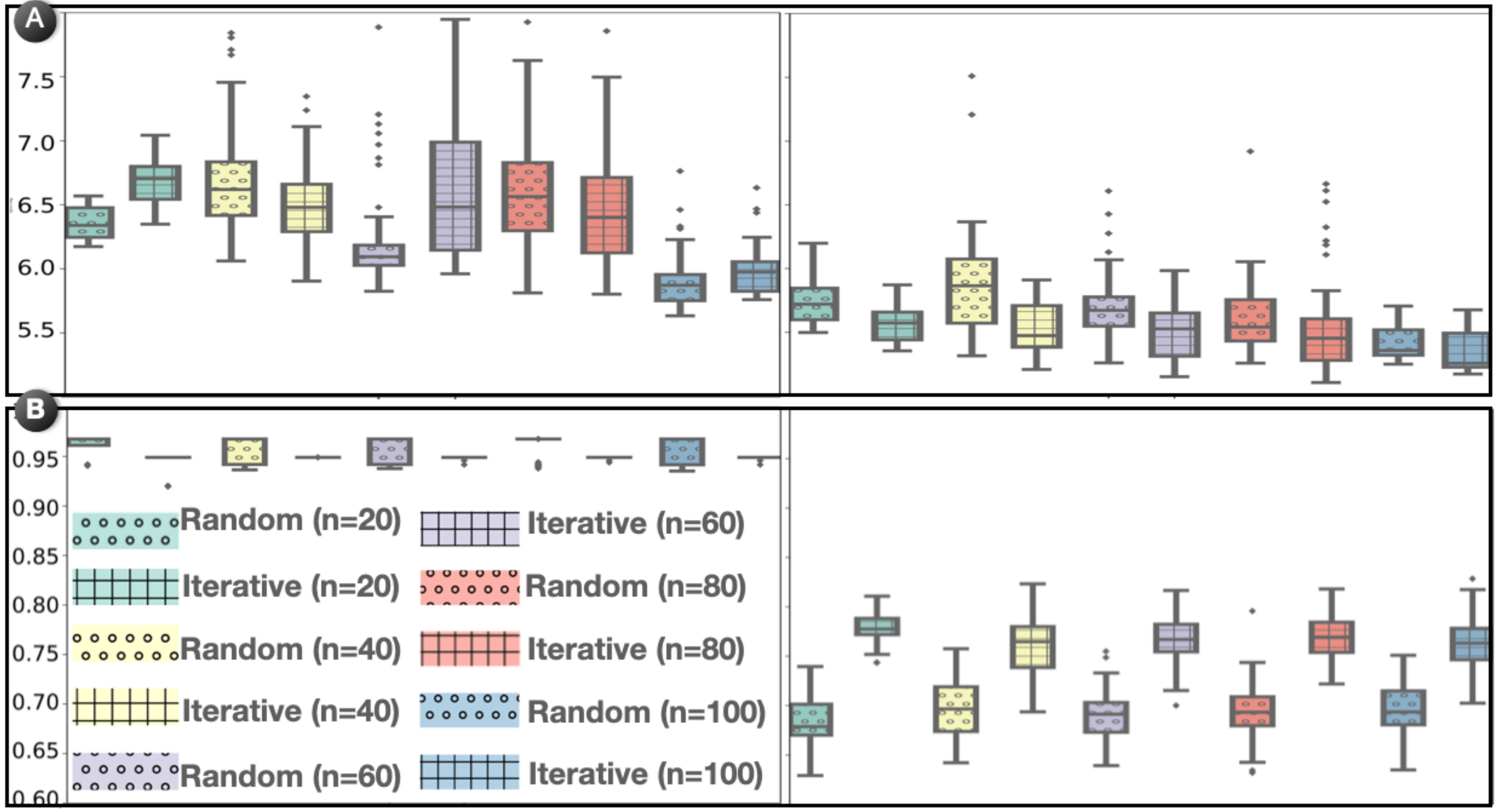}
    \vspace{-0.25in}
    \caption{The (A) running time and (B) Adjusted Mutual Information (AMI) score of tSNE (left) and UMAP (right) with different parameters settings.}
    \label{fig:evaluation_iterative}
\end{figure}
\setlength{\belowcaptionskip}{0pt}

The performance of \textit{hard recommendation} can be revealed from the successful loss converging pattern and high prediction accuracy. 
To train the model, we utilize a dataset consisting of 1591 survey questions with target variables as labels, which is manually labeled by the domain experts during the pre-processing of harmonization. The dataset is split into 90\% training and 10\% validation. 
The training hyper-parameters are set as follows: batch-size=32, number-of-epochs=10, learning-rate=2e-5. The model converging process is depicted by the loss shown in \autoref{fig:loss}A. The validation accuracy reaches 99\% in the final epoch, which promises good performance on the classification task. 

Even with the high performance of \textit{hard recommendation}, some user-defined text is likely related to multiple target variables where \textit{soft recommendation} is more applicable. \textit{Soft commendation} promises to capture the semantic similarity in the embedding projection space so that users are able to identify multiple related survey questions. Therefore, the clusters formed in the embedding space should be verified to capture the semantic information. We decided to compute the similarity between the clustering of survey question embeddings and the groups of ground truth, expecting those questions with the same labels will be grouped into the same cluster in the projection space. To demonstrate the good performance of the embedding clustering results, we compare the embeddings generated from our fine-tuned model with the pre-trained model. Besides the embedding representation, the performance also relies on projection methods. Thus, we also take two projection methods into consideration, i.e. tSNE~\cite{van2008visualizing} and UMAP~\cite{mcinnes2018umap}.

We choose Adjusted Mutual Information (AMI) to quantitatively compare the clusters from embedding and ground truth labels. AMI is a widely used method to compare different partition/clustering results of the same dataset~\cite{vinh2010information}. When comparing 
two clustering results, 
since the mutual information (MI) for the result with  a larger number of clusters is generally higher, AMI takes this into account and adjusts MI through the following equation:
\begin{equation}\small
AMI(U,V) = \frac{MI(U,V)-E\{MI(U,V)\}}{max\{H(U), H(V)\}-E\{MI(U,V)\}}    
\end{equation}where U, V are the results of two clustering methods. The value of AMI is in the range of [0,1], where 1 means U and V are identical.

As shown in \autoref{fig:loss}B, color is used to differentiate fine-tuned  (\tikzcircle[fill={rgb,255:red,142; green,211; blue,199}]{3pt}) and pre-trained model (\tikzcircle[fill={rgb,255:red,255; green,255; blue,179}]{3pt}). Within each group, the projection method is encoded in textures of the boxplot. It is clear to see that fine-tuned model improves the results a lot for both tSNE and UMAP because the  results of embedding clustering from the fine-tuned model much better match the ground truth. Also, we can observe that tSNE outperforms UMAP in both models. We can conclude that the fine-tuning improves the \textit{soft recommendation} regardless of the projection methods. 
% \textit{Adjusted measures} are common ways to compare different partition/clustering results of the same dataset, and AMI is suitable when \clrb{the ground truth clustering is unbalanced.}
% We choose Adjusted Mutual Information (AMI) to quantitatively prove the fine-tuned model outperforms the pre-trained model in capturing the semantic meaning. 
% \clr{the numbers of clusters from two partition/clustering results are very different (please confirm if this description is correct or not)}
%  Using AMI, we quantitatively evaluate the projection results of UMAP and tSNE for both the pre-trained and fine-turned BERT models. 
% 

\subsubsection{Qualitative Evaluation}

% To demonstrate the improvements of projection space, we compare the question embeddings in the scatterplots of \sysname{} for two models:  fine-tuned BERT classification model, pre-trained BERT model as baseline model \clr{ref}.
To demonstrate the qualitative improvements of the fine-tuned model, \textit{Scatterplots} with the two clustering results from different models are shown in \autoref{fig:evalution_embedding}. The figure discloses several advantages of our fine-tuned model ($a_{2}$) compared with the pre-trained BERT ($a_{1}$). First, there is a clear boundary between different clusters in \autoref{fig:evalution_embedding}-$a_{2}$. While in the pre-trained model, several groups interfere with each other, and it is hard to differentiate them without coloring. Second, related groups are also closer to each other in \autoref{fig:evalution_embedding}-$a_{2}$, which corresponds to high-level sociology concepts. As pointed by the experts, \say{trust in people}, \say{trust in party}, \say{trust in government}, and  \say{trust in legal system} form the concept of \say{trust in political institutions}. \say{interest in politics} is close to \say{never discuss politics}, both of them depict the attitude of respondents in politics. \say{demonstration} and \say{signing petition} comprise the concept of \say{political behaviors}.

We would also like to give an example to compare the quality of \textit{soft recommendation}. As shown in \autoref{fig:evalution_embedding}, \faPlus~ indicates the projected embedding of a user's input, i.e. \say{trust in parliament}. 
From the result of pre-trained model (\autoref{fig:evalution_embedding}-$a_{1}$), it is clear that the queried topic is isolated from  multiple topics in the projection. After brushing the surrounding circles, the table presents some non-related target variables. However, in our fine-tuned BERT model (\autoref{fig:evalution_embedding}-$a_{2}$), the queried topic falls into a small cluster of circles. The cluster (i.e., trust in parliament) presents related target variables (\texttt{T\_TRPARL\_11, T\_TRPARL\_DISTRIB}) for the queried topic. 
We can conclude that the training not only teaches the model to better predict target variables but also significantly improves the performance of \textit{soft recommendation}. 

\setlength{\textfloatsep}{0.1cm}
\begin{table}[]\small
\centering
\begin{tabular}{|c|c|c|c|}
\hline
\multicolumn{2}{|c|}{\cellcolor{gray!25}Concept}                                                                                                                                                 & \cellcolor{gray!25}Name                & \cellcolor{gray!25}Label                                                                     \\ \hline
\multirow{5}{*}{\begin{tabular}[c]{@{}c@{}}political\\ attitudes\end{tabular}} & \multirow{4}{*}{\begin{tabular}[c]{@{}c@{}}trust in\\ political\\ institutions\end{tabular}} & T\_TRPARL\_DISTRIB  & trust in the parliament                                                   \\ \cline{3-4} 
                                                                               &                                                                                              & \texttt{T\_TRLEG\_DISTRIB}   & trust in the legal system                                                 \\ \cline{3-4} 
                                                                               &                                                                                              & \texttt{T\_TRPARTY\_DISTRIB} & trust in political parties                                                \\ \cline{3-4} 
                                                                               &                                                                                              & \texttt{T\_TRGOV\_DISTRIB}   & trust in the government                                                   \\ \cline{2-4} 
                                                                               & interests                                                                                    & \texttt{T\_INTPOL\_DISTRIB}  & interest in politics                                                      \\ \hline
\multicolumn{2}{|c|}{\multirow{2}{*}{political behavior}}                                                                                                                     & \texttt{T\_DEMONST}          & \begin{tabular}[c]{@{}c@{}}participation in\\ demonstrations\end{tabular} \\ \cline{3-4} 
\multicolumn{2}{|c|}{}                                                                                                                                                        & \texttt{T\_PENTITION}        & signing petitions                                                         \\ \hline
\multicolumn{2}{|c|}{\multirow{4}{*}{Socio-demographics}}                                                                                                                     & \texttt{T\_AGE}              & age                                                                       \\ \cline{3-4} 
\multicolumn{2}{|c|}{}                                                                                                                                                        & \texttt{T\_GENDER}           & gender                                                                    \\ \cline{3-4} 
\multicolumn{2}{|c|}{}                                                                                                                                                        & \texttt{T\_METRO}            & living in metropolitan                                                    \\ \cline{3-4} 
\multicolumn{2}{|c|}{}                                                                                                                                                        & \texttt{T\_EDU}              & education                                                                 \\ \hline
\end{tabular}
\caption{Expert-defined sociology concepts, corresponding target variables and labels.}
  \label{TAB:CONCEPT}
\end{table}

\subsection{Evaluation of Embedding Iterative Updating Algorithm}
This section measures to what extent our \textit{Embedding Iterative Updating Algorithm} can stabilize the projection results, and how much it can improve the projection efficiency. The study was conducted by running the algorithm with different iterations (i.e., 20, 40, 60, 80, and 100).
% We run the \textit{Embedding Iterative Updating Algorithm} with iteration=20,40,60,80, and 100. 
To show the effectiveness of our algorithm, a baseline of updating the embedding projections with random initialization was also conducted. 
% We compare the running time and AMI scores under different parameter settings. 

We compute both the running time (\autoref{fig:evaluation_iterative}A) and AMI score (\autoref{fig:evaluation_iterative}B) of tSNE (left) and UMAP (right) with different parameter settings. The $x$-axis indicates the increasing number of iterations utilized in our iterative updating algorithm.
The \textit{random initialization} and \textit{iterative updating} are encoded in the texture.
Comparing to UMAP (right), tSNE (left) takes longer time, but generates better clustering results. 
When it comes to the efficiency between \textit{random initialization} and \textit{iterative updating},  tSNE cannot guarantee \textit{iterative updating} will help the algorithm converge faster (upper left). But it is clear to see that \textit{iterative updating} decreases the running time for UMAP (upper right).
For the AMI score, with tSNE projection, \textit{iterative updating} does not improve accuracy given the fact that \textit{random initialization} already reaches a high-level (bottom left). But we can also infer from the figure that \textit{iterative updating} makes the iteration stable by decreasing the variance of running time. For UMAP, \textit{iterative updating} can improve the clustering results regardless of the running iterations (bottom right). 
\begin{figure}
    \centering
    \includegraphics[width=\linewidth]{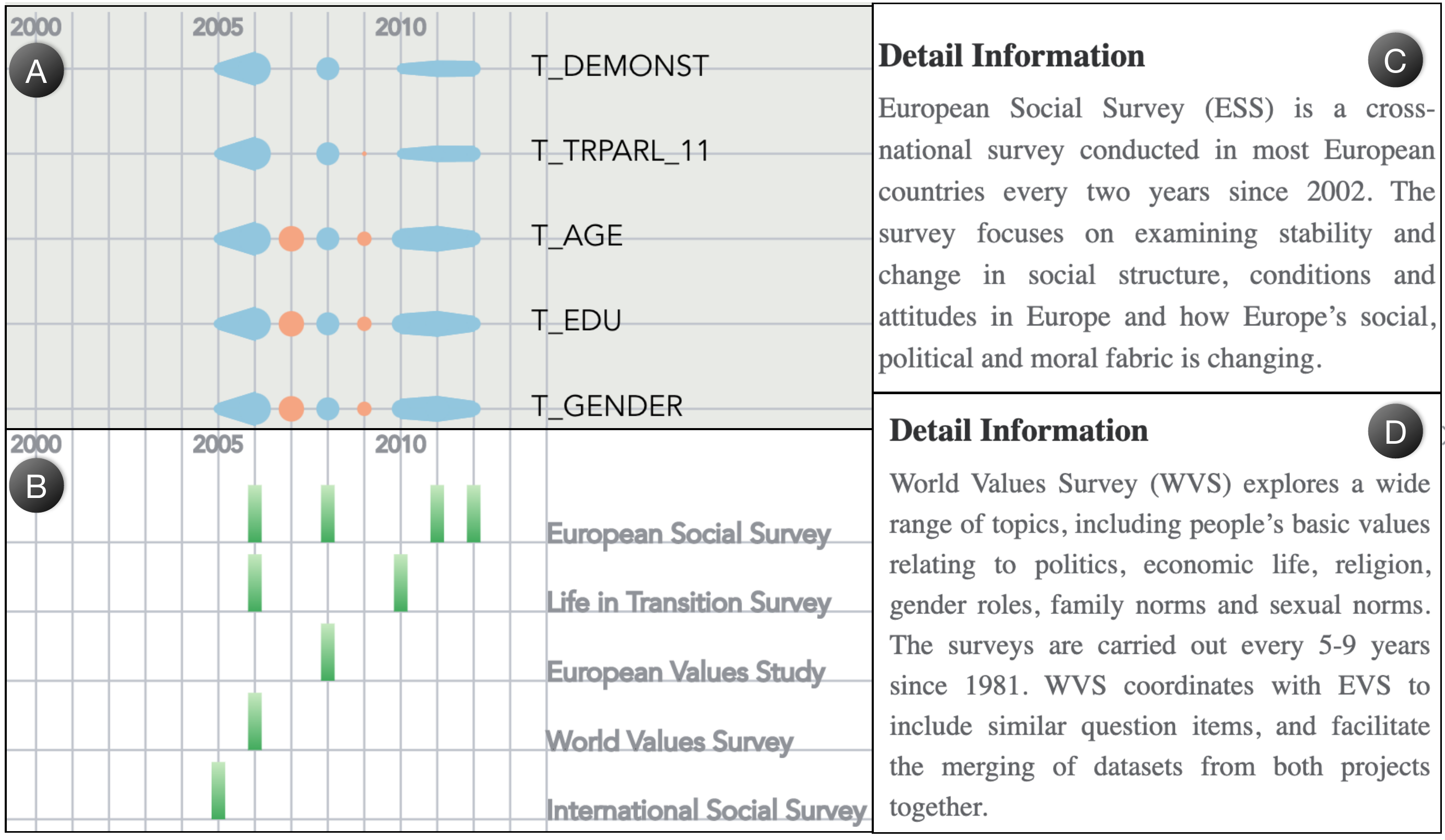}
    \vspace{-0.25in}
    \caption{The data availability for analyzing protest participation in Russia.}
    \label{fig:Russian}
\end{figure}

\vspace{-1em}
\subsection{Data Availability Checking}
We illustrate how the \newdesign{} can be utilized in different scenarios with two case studies: one emphasizes how it can help the SDR research group to summarize the possible directions for social science research; the other describes how it can help scientists to decide whether data are sufficient or not for comparative analysis. All the names used in the case studies are pseudonyms for privacy issues. 

\subsubsection{Case1: Political engagement: attitudes and behaviors}
We invite an expert, Arya, who works on the SDR project and has a deep understanding of the harmonization process of SDR. Arya wants to propose future research directions/topics according to available data of the SDR regarding political engagement. 

Since Arya knows the variables very well, she jumps to the QBC directly to check the data availability. 
First, Arya selected some target variables to form the high-level theoretical concepts, summarized in \autoref{TAB:CONCEPT}. Without adding filtering conditions, Arya clicked the \faSearch~ button, the availability of selected variables is displayed in \autoref{fig:teaser} ($b_{1}-b_{3}$)

\setlength{\dbltextfloatsep}{-6pt}
\begin{table*}[]\small
\begin{tabular}{|l|l|l|}
\hline
\cellcolor{gray!25}Survey                                                                                                 & \cellcolor{gray!25}Background                                                                                                                                                                                                                                                                                                                              & \cellcolor{gray!25}Insights                                                                                                                                                                                                                                                                                                                                        \\ \hline
\begin{tabular}[c]{@{}l@{}}European Social \\ Survey (ESS)\end{tabular}                                & \begin{tabular}[c]{@{}l@{}}ESS  aims to examine stability and change in social  structure, \\ conditions, and attitudes in Europe, which is conducted in most \\ European countries since 2002.\end{tabular}                                                                                                                      & \begin{tabular}[c]{@{}l@{}}Jimmy noticed that the available surveys contain several well-known \\ and widely used surveys. ESS is one of the examples, shown in \autoref{fig:Russian}C\end{tabular}                                                                                                                                                            \\ \hline
\begin{tabular}[c]{@{}l@{}}International Social \\ Survey Program (ISSP)\end{tabular}                  & \begin{tabular}[c]{@{}l@{}}This survey is a continuous program of cross-national collaboration \\ running surveys, covering multiple issues related to social structure.\end{tabular}                                                                                                                                                   & \begin{tabular}[c]{@{}l@{}}Some of them are cross-national collaboration surveys while others are \\ conducted in specific regions.\end{tabular}                                                                                                                                                                                                \\ \hline
\begin{tabular}[c]{@{}l@{}}Life in Transition \\ Surveys (LITS)\end{tabular}                           & \begin{tabular}[c]{@{}l@{}}LITS was carried out by the European Bank for Reconstruction and \\ Development in collaboration with the World Bank in 2006 and 2010 in \\ central-eastern Europe and the Baltic States, south-eastern Europe, etc.\end{tabular}                                                                            & \begin{tabular}[c]{@{}l@{}}Jimmy  learned  that  some  unavailability  comes from the surveys, \\ not the selected target variables. The reason is that they were not \\ conducted as continuous programs. For example, LITS was\\ conducted only in 2006 and 2010.\end{tabular}                                                                                                                   \\ \hline
\begin{tabular}[c]{@{}l@{}}World Values \\ Survey (WVS) \\ European Values \\ Study (EVS)\end{tabular} & \begin{tabular}[c]{@{}l@{}}WVS focused on a wide range of topics, including economic life, \\ religion, basic values relating to politics. EVS are conducted every 9 \\ years in most European countries since 1981, which examines social, \\ political, and economic values and attitudes, as well as living conditions.\end{tabular} & \begin{tabular}[c]{@{}l@{}}Jimmy identified some relationships between surveys from the background \\ information. WVS can be merged with EVS and used together since they \\ share similar survey questions. From the Separate Availability in \autoref{fig:Russian}B, \\ both cover different years, facilitating extensive evolution analysis.\end{tabular} \\ \hline
\end{tabular}
\caption{Descriptions and insights about available surveys.}
  \label{TAB:ASNETWORK}
\end{table*}

From the color (\tikzcircle[fill={rgb,255:red,146; green,197; blue,222}]{3pt}) in \autoref{fig:teaser}-$b_{1}$, the joint available data for all the selected variables are available for years 1990, 1995-1998, 2004-2011. Arya checked each concept separately. In the concept of socio-demographics, the data for \say{age} and \say{gender} are pretty complete. However, there are no sufficient data for \say{living in metropolitan} in the 70s and in the beginning of the 80s. Also, \say{education} has a deficiency gap during 1982-1984. Given the incomplete socio-demographics of respondents, Arya concluded that those years with data deficiency should be imputed or excluded by  researchers.
For concept of political attitudes,  \say{interest in politics} has the greatest data coverage. 
Based on the observation that \say{trust in the parliament} and \say{trust in the legal system} share the same temporal coverage, Arya confirmed that it allows researchers to conduct a study about trust in political institutions from the 80s, even with the gap from 1985-1988. Compared to the \say{political attitudes}, \say{political behaviors} can be analyzed more comprehensively from the 60s given the higher temporal coverage.

Drilling down to the country coverage, Arya clicked several available surveys in the \textit{Joint Availability} to check it. The result is presented in \autoref{fig:teaser}-b3, it is clear to see that the available data (\tikzcircle[fill={rgb,255:red,204; green,235; blue,197}]{3pt}) cover Latin America (Latinobarometro), Europe (World Values Study), and Asia (Asia Europe Survey). She concluded that researchers can conduct various analyses on political attitudes and behaviors controlling for socio-demographic characteristics for a period of over 40 years in different regions.  
She also pointed out that even in the same survey project conducted in different years, the covered region can fluctuate. The beauty of the SDR harmonized data is that different surveys can complement each other not only  temporally but also spatially. For example, when Arya hovered over the bar chart, it showed that World Values Survey (WVS) covers 23 countries in 2006. From the detailed coverage map visualized in \autoref{fig:teaser}-$b_{3}$, it is obvious to see the covered area  
includes Latin America. However, it only covers 9 countries in 2007 without Latin America, which can be supplemented by another survey project in 2007, i.e., Latinobarometro.
Finally, Arya summarized that regarding political engagement, the available survey projects offer a substantive set of variables for  comparative cross-national research. %, as well as for a longitudinal study.
% \begin{figure}
%     \centering
%     \includegraphics[width=\linewidth]{figure/concept_case.pdf}
%     \caption{TBA}
%     \label{fig:case1}
% \end{figure}
\vspace{-0.6em}
\subsubsection{Case2: Protest participation in Russia}

Given the possible research directions of the SDR harmonized dataset, Jimmy wants to study protest participation in autocratic states by analyzing contemporary Russia. He needs data that cover Russia in the 2000s$\sim$2010s, the period of the tightening autocratic measures in the country. 
Based on the literature review, Jimmy proposes that several determinants, such as trust in political institutions, satisfaction with the democratic performance in the country, and economic hardship, can influence protest participation in autocracies in different ways.

After exploring the system with QBQ, he decided to use \texttt{T\_DEMONST} as a protest indicator and \texttt{T\_TRPARL} as an indicator for trust in political institutions. 
He also identified a set of necessary socio-demographic variables for his study. However, the SDR data lack  two variables that measure potential protest determinants, theorized by Jimmy based on the literature review, namely, the subjective perception of democracy and individual’s economic situation. 
After detecting variables in the SDR data, Jimmy applied two filtering conditions to check the data availability via QBC: (country==\say{Russian}, year${\in}$[2000, 2019]).

While examining the joint availability displayed on \autoref{fig:Russian}A, 
Jimmy identified that the unavailability gaps in 2007 and 2009 come from the lack of \texttt{T\_TRPARL\_11} and \texttt{T\_DEMONST}. He concluded that the sufficient for his study data on Russia are available only from 2005 to 2012 with gaps in \say{demonstration} and in ``trust in parliament''. Therefore, he needs to make the decision whether to use biannual data (i.e., data from year 2006, 2008, 2010, 2012), impute the missing data from 2007 to 2009 or search for another dataset because of the missing data.
Also, it is clear to see that the size of available samples fluctuates over time, which might be a concern regarding how to use the data properly, as pointed out by Jimmy. The information derived from our visualization helps Jimmy to structure his research, and the prior-knowledge on the data availability leads to more reasonable expectations on the final outcome. Drilling down to the \textit{Joint Availability} in \autoref{fig:Russian}B,  several findings and derived insights are summarized in \autoref{TAB:ASNETWORK}.

Based on these observations and conclusions, Jimmy agreed that the combination of these surveys ensures sufficient high-quality samples. However, Jimmy’s main concern is the lack of key variables, i.e. two potential determinants of protest participation. 
He concluded that using only SDR data is not
 sufficient for his research due to the time and variable coverage limitations.
He probably can try to harmonize the data for missing variables by himself from other sources. 
% \textit{Trust in Parliament} 
% Based on literature survey, the 

\subsection{Participation in Demonstrations worldwide}

We invite expert Kiara to demonstrate how \sysname{} can assist scientists in the social research process following one previous study~\cite{kolczynska2020micro}.
It is focused on participation in demonstrations worldwide, which requires samples with a high regional diversity for comparison. 
Kiara hypothesized that resources and political attitudes have different effects on the levels of participation in demonstrations in democratic and non-democratic countries. 
Kiara was curious if she could rely on the SDR data. Before downloading the dataset to conduct further analysis, she used our \sysname{} to check the data.
% \begin{enumerate}[leftmargin=0.18in, topsep=-0.2em]
% 	\itemsep-0.25em
%   \item (Understanding) whether the data had variables of interest to conduct her research, including both the indicators of demonstrations and determinants from her theory.
% %   and whether there were other variables that she needed to consider for her analysis.
%   \item (Exploring) whether there were enough data with high quality, large regional diversity and continuous temporal coverage. 
%   \item (Evaluating) whether the available variables she can include in the regression model were correlated with each other.  
% \end{enumerate}

To begin with, Kiara utilized QBQ to explore the variables that she could use. Kiara first typed in the most important concept of her research, \say{participation in demonstration}. 
The prediction from \say{hard recommendation} is \say{\texttt{T\_DEMONST}} in \autoref{fig:teaser}-$a_{3}$, which conforms to her domain knowledge. 
Kiara knew that the meaning of demonstration can vary a lot depending on political backgrounds, thus, she was wondering what is the definition of \say{\texttt{T\_DEMONST}} in the SDR. 
From \autoref{fig:teaser}-$a_{1}$, it is clear that \faPlus~ falls into one cluster, from which Kiara brushed circles (\tikzcircle[fill={rgb,255:red,253; green,191; blue,111}]{3pt}) to check the detailed information. 
The table in \autoref{fig:teaser}-$a_{2}$ showed that the source variables for \say{\texttt{T\_DEMONST}} include not only demonstrations, but also protests and marches. It also varied between participation in a public, authorized, or unauthorized demonstrations, as well as very specific protests (e.g., against the former president). 
The source questions also varied in terms of the time length. Generally, respondents were asked in the format of \textit{\say{Have you performed [action type] in the last [time period]?}}, where [time period] varied across \say{twelve months}, 1, 2, 3, 4, 5, 8, 10 years or ever in different surveys.   

 Kiara intended to learn the structure of  harmonized data via \textit{circular graph}. From there, many visual investigations can be done through interactive exploration. 
With a brief overview, Kiara found that a four-level hierarchical structure correspond to the type of variables: socio-demographics, quality control variables, target variables, harmonization control variables, which differs from the common one-survey data structure. 
She hovered the highlighted bar to see the predicted variable's name, i.e. \say{\texttt{T\_DEMONST}}. It faces 
four arcs, which indicates that variations in source questions were captured with four harmonization control variables. 
She was inquisitive about the meaning of control variables, so she clicked one of them, i.e., \texttt{C\_PR\_DEMONST\_YEARS}.
The labels and distributions are shown in \autoref{fig:teaser}-$a_{4}$, she discovered the variance of the time range is captured well and the most frequent asked year range is \say{ever}. 
After the in-depth exploration, Kiara confirmed that the information contained in control variables revealed the high quality of the harmonized data and allowed her to conceptualize participation in demonstrations for her study.

After the preliminary examination of the available variables, Kiara decided to choose \texttt{T\_DEMONST} as the indicator for participation in demonstrations. When deciding on the measurement for political attitudes, she found a variety of options. 
During our tutorial session, Kiara learned that the length of each target variable bar indicates its popularity in different surveys and countries. Thus, she selected \texttt{T\_TRPARL} (trust in parliament) to measure political attitudes based on the distributions for all the political attitude variables.  
To measure resources, she picked \texttt{T\_EDU} (education). She selected \texttt{T\_GENDER} and \texttt{T\_AGE} to control for the socio-demographic characteristics of the respondents from the sample. Kiara proceeded further with this set of variables. 

\begin{figure*}[ht]
    \centering
    \includegraphics[width=\textwidth]{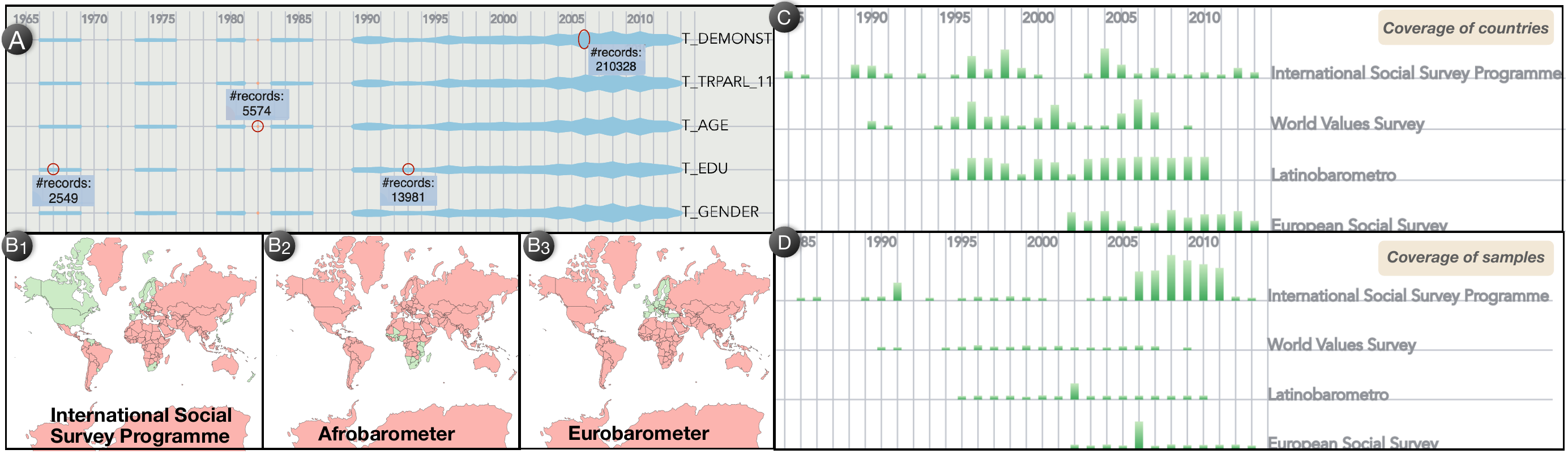}
    \vspace{-0.25in}
    \caption{ Case study: Participation in demonstrations worldwide (A) \textit{Separate Availability} (B1-B3) country coverage of several survey projects; (C) top-3 survey projects with highest availability (macro-level); (D) (micro-level)}
    \label{fig:Marta}
\end{figure*}

From the \textit{Separate Availability} in \autoref{fig:Marta}A, Kiara easily identified the possible time-period of her study, as SDR provides sufficient data from 1989 to 2013 without any gap. The gap in 1982 can be explained by the deficiency of  \texttt{T\_EDU}. 
From the \textit{responsive bar charts}, there are 22 available surveys in total, which convinced Kiara that the increase in data coverage is one of the primary advantages of the harmonized data. 
We displayed the top-3 surveys with highest availability in \autoref{fig:Marta}C (macro-level) and \autoref{fig:Marta}D (micro-level).
The pattern indicates some surveys have stable country coverage, e.g., Latinbarometro. While other surveys fluctuate a lot, e.g., International Social Survey Programme. 
Kiara found that the available data also cover diverse regions and countries in \autoref{fig:Marta} (B1-B3), which allows her to compare democratic and non-democratic countries from different parts of the world. 
\setlength{\belowcaptionskip}{5pt}
\setlength{\belowcaptionskip}{5pt}

Next, Kiara wanted to query the data patterns to verify if selected variables are correlated with each other, as this was an important condition for including variables in her regression analysis. 
In the correlation matrix(\autoref{fig:teaser}-$c_{1}$), the intense color of all cells indicated that the correlations between all the variables were significant. Based on this observation, Kiara concluded that the variables selected in her model were accurate and can be used in her model. Finally, Kiara clicked the cell in the correlation matrix to look deeper into the relationship between \texttt{T\_DEMONST} and \texttt{T\_EDU}. She then found out that these two target variables are correlated not only with each other but also with their respected harmonization control variables and  with quality control variables (\autoref{fig:teaser}-$c_{2}$). 
Furthermore, based on the significance level labeled on each edge, she also identified that the \texttt{T\_EDU} has weaker relationships with quality control variables compared to \texttt{T\_DEMONST}, indicating \texttt{T\_EDU} has  better quality than \texttt{T\_DEMONST}.
In conclusion,  Kiara verified that she would include those methodological variables (i.e., quality- and harmonization control variables) in her regression analyses as well. 

After such detailed exploration, Kiara concluded that the ex-post harmonized survey data are sufficient for her research. The country and time coverage allowed her to study the effect of education and trust in parliament on the probability of individuals to participate in demonstrations in democratic and non-democratic countries. She also pointed out that SDR lack macro-level data with democracy indicators, which she needs to add from the other dataset. At the micro-level, SDR data contain all the items she needs. Besides, the interface demonstrates the high quality of the SDR data and the importance of including methodological variables while using a harmonized dataset, such as SDR. She agreed that \sysname{} provides accurate guidance for variables identification, supplies efficient decision-making support for relying on the harmonization data or not via visual exploration and contributes a lot for variable selection in regression models.

% The embedding clustering is not only useful in our automatic variable recommendation, but also can be utilized to efficiently harmonize related source variables in pre-processing stage. In this section, 

\section{Expert Feedback}

We conducted in-depth interviews with the same group of experts to gather their qualitative feedbacks on the usefulness and usability of \sysname{} (\textit{E1${\sim}$E4}). E1 and E2 are social scientists with more than 45 years of experience studying social movements and contentious politics. E3 has 10+ years of experience in survey data harmonization and E4 has 5+ years of experience in both survey data transformation and data management. 
We started the interviews with an introduction of the \sysname{} pipeline and the functions of individual visual components. 
The interviews were in an interactive way to discuss the pros and cons, suggestions, and  agreements on \sysname{}. 

Overall, all experts agreed that the tool  \say{ is very helpful in learning the structure and capacity of the harmonized survey data}, and \say{ also contributes to data methodology literature, for proposing new ways to work with the harmonized dataset.}  
As to the goal of \sysname{}, E1 commented that \say{given the ambitious goal, it can bring a big contribution to social science as a pioneer research}. E2 added that \say{\sysname{} will increase the popularity of the SDR given the novel visualizations, making it available to researches from other fields using survey data for analysis, such as economy or psychology.}
In terms of the usability, they agreed that it is easy to understand the system without knowledge in computer science and visualization.
Regarding the assistance for scientists to identify desired data, they believed \say{the availability checking is extremely needed given the complexity of the SDR data} and \say{the visual interactions are indeed useful to explore the complex data}.

All experts expressed their interest in the QBQ component, E4 mentioned that \say{the role of the module will be even bigger when the number of variables increases as the harmonized data become  mature in the future.} E4 also pointed out that \say{besides the ex-post harmonization analysis, the trained model is also useful for social scientists to retrieve questions during the harmonization process.} E3 agreed and summarized that \say{QBQ can be used in three different levels. Besides ex-post analysis and pre-processing  during harmonization, it can be extremely useful for international surveys as they contain hundreds of variables. Although social scientists do not quite understand the inner structures of the trained model, it is efficient to identify related variables and easy to use.}
E2 evaluated QBR highly, as harmonized data users should consider methodological variables to include in their models, and added that \say{it is very useful to present the network to scientists given the complex relationships among the variables.} E1 highlighted the importance of the visualising some weak relations between target variables and quality control variables, as it indicates good quality of the samples. 

Additionally, the experts suggested some improvements for \sysname{}. E1 expressed a concern that users might overthink the meaning of each element after learning visual mappings introduced in \sysname{}. Take the network in QBR as an example, after noticing the color of edge indicating the coefficient is valid or not, they may wonder whether the edge length also encodes other information. Both E3 and E2 were first confused about the coloring schema of the  \newdesign{}, though they understood it after detailed explanations. They worried that the learning curve of the coloring algorithm might be high for social scientists without visualization and database training. %, the learning curve of the coloring algorithm might be high for a minority of them. %
These comments will be considered in the  future when deploying \sysname{} into the SDR portal.

\section{Conclusion}
In this work, we present \sysname{}, a visual query system that facilitates scientists to locate target data and evaluate their theoretical models. To achieve this, the system provides visual guidance and queries with three modules: Query-by-Question, Query-by-Condition, Query-by-Relation. From the solid evaluation and thorough studies, we have identified several applications for QBQ and exemplified how QBC and QBR help scientists to understand, explore, and utilize harmonized survey data in their research. Insightful findings and positive feedback from domain experts demonstrated the novelty, usefulness and effectiveness of \sysname{}.

\bibliographystyle{abbrv-doi}
\bibliography{template}
\end{document}